\documentclass[11pt,a4paper]{article}
\usepackage{amssymb}
\usepackage{amsmath}\usepackage{graphicx}  
\numberwithin{equation}{section}
\usepackage{hyperref}
\usepackage{url}
\usepackage{footnote}
\makesavenoteenv{tabular}
\newcommand{\p}{\partial}

\newcommand{\id}[1]{\ensuremath{\mathrm{id}}}

\newcommand{\half}{\mbox{\footnotesize $\frac{1}{2}$}}

\newcommand{\qm}{quantum mechanics}
\newcommand{\er}{\eqref}
\renewcommand{\L}{\label} 
\newcommand{\beq}{\begin{equation}}
\newcommand{\eeq}{\end{equation}} 
\newcommand{\bea}{\begin{eqnarray}}
\newcommand{\eea}{\end{eqnarray}}

\newcommand{\ovl}{\overline}
\newcommand{\ul}{\underline}

\newcommand{\raw}{\rightarrow}

\newcommand{\Hs}{Hilbert space} 
 
 \newcommand{\cci}{C^{\infty}_c}

\newcommand{\hv}{hidden variable}

\newcommand{\al}{\alpha} 
 
 \newcommand{\Dl}{\Delta}

\newcommand{\lm}{\lambda} \newcommand{\Lm}{\Lambda}
\newcommand{\rh}{\rho} \newcommand{\sg}{\sigma}

\newcommand{\ch}{\ch} 
\newcommand{\ps}{\psi}
\newcommand{\om}{\omega} 

\renewcommand{\L}{\label}

\newcommand{\N}{{\mathbb N}} \newcommand{\R}{{\mathbb R}}
 
  \makeatletter
 
\makeatletter
\def\moverlay{\mathpalette\mov@rlay}
\def\mov@rlay#1#2{\leavevmode\vtop{%
   \baselineskip\z@skip \lineskiplimit-\maxdimen
   \ialign{\hfil$\m@th#1##$\hfil\cr#2\crcr}}}
\newcommand{\charfusion}[3][\mathord]{
    #1{\ifx#1\mathop\vphantom{#2}\fi
        \mathpalette\mov@rlay{#2\cr#3}
      }
    \ifx#1\mathop\expandafter\displaylimits\fi}
\makeatother

\newtheorem{definition}{definition}%[section]

\newtheorem{theorem}[definition]{Theorem}

\topmargin = - 1 cm \textheight = 23 cm \textwidth = 15 cm
\usepackage[symbol]{footmisc}
\renewcommand{\thefootnote}{\fnsymbol{footnote}}
\begin{document}
\pagenumbering{arabic} \setlength{\unitlength}{1cm}\cleardoublepage
\date\nodate
\begin{center}
\begin{huge}
{\bf Bohmian mechanics is not deterministic}\footnote[1]{Submitted to the special issue of  \emph{Foundations of Physics}: `The pilot-wave and beyond:  Celebrating Louis de Broglie's and David Bohm's quest for a quantum ontology'. The claims in this paper  follow from the general analysis of randomness in \qm\ in Landsman (2017, 2020, 2021) and were first presented at a Harvard Foundations of Physics seminar (organized by Jacob Barandes), December 15, 2020, in a talk called `Determinism and undecidability:
You can't eat your cake and have it too.' I am grateful to the audience for several critical questions, which have clarified my reasoning. Further improvements were made after a discussion at the Cambridge--LSE Philosophy of Physics Bootcamp, May 24, 2022. Finally, I am indebted to two anonymous referees for numerous constructive suggestions, which I took into account.
}
\end{huge}
\bigskip\bigskip

\begin{Large}
 Klaas Landsman\vspace{5mm}
 \end{Large}
 
 \begin{large}
 Department of Mathematics, 
Institute for Mathematics, Astrophysics, and \\ Particle Physics (IMAPP), Radboud University, Nijmegen, The Netherlands
Email:
\texttt{landsman@math.ru.nl}
\end{large}
 \begin{abstract} 
\noindent 
I argue that Bohmian mechanics (or any similar pilot-wave theory) cannot reasonably be claimed to be a deterministic theory. If one assumes the ``quantum equilibrium distribution'' provided by the wave function of the universe,
 Bohmian mechanics requires an external random oracle 
in order to describe the (Kolmogorov--Levin--Chaitin) algorithmic randomness properties of typical outcome sequences of long runs of repeated identical experiments (which provably follow from the Born rule). This oracle lies beyond the scope of Bohmian mechanics (or any deterministic extension thereof), including the impossibility of explaining the randomness property in question from ``random'' initial conditions. 
 Thus the advantages of Bohmian mechanics over other interpretations of \qm, if any, must lie at an ontological level, and in its potential  to derive the quantum equilibrium distribution and hence the Born rule. 
\end{abstract}\end{center}

\thispagestyle{empty}
\renewcommand{\thefootnote}{\arabic{footnote}}
\section{Introduction}\label{intro}
Bohmian mechanics is widely seen as a \emph{deterministic} interpretation of \qm. This point was already made by Bohm himself, and has been repeated by his followers:\footnote{See Kaiser (2011), Norsen (2017),  Greenstein (2019), Bricmont (2022), Bub (2022), and Ryckman (2022) for some history of Bohmian mechanics and the closely related de Broglie--Bohm pilot wave theory. }
\begin{quote}
\begin{small}
My interpretation of the quantum theory describes all processes as basically causal and continuous.\footnote{
Here `causal' is  synonymous to `deterministic'. This is clear from a contrast Bohm draws on the same page: `Thus, we are able in my interpretation to understand by means of a causal and continuous model just those properties of matter and light which seem most convincingly to require the assumption of discontinuity and incomplete determinism.' Similarly, Holland's well-known book \emph{The Quantum Theory of Motion} is  subtitled: \emph{An Account of the de Broglie--Bohm Causal Interpretation of Quantum Mechanics}.
}
 (Bohm, 1952, p.\ 175)
\end{small} 
\end{quote}
\begin{quote}
\begin{small}
But in 1952 I saw the impossible done. It was in papers by David Bohm. Bohm showed explicitly how parameters could indeed be introduced, into nonrelativistic wave mechanics, with the help of which the indeterministic description could be transformed into a deterministic one.
 (Bell, 1982, p.\ 990)
\end{small} 
\end{quote}

\begin{quote}
\begin{small}
Bohmian mechanics happens to be deterministic. A substantial success of Bohmian mechanics is the explanation of quantum randomness or Born's statistical law, on the basis of Boltzmann's principles of statistical mechanics, i.e.\ Born's law is not an axiom but a theorem in Bohmian mechanics. (D\"{u}rr \& Teufel, 2009, p.\ 6)
\end{small} 
\end{quote}

My aim is to show that the alleged determinism of Bohmian mechanics is parasitical on some external random sampling mechanism (``oracle'') the theory has to invoke in order to state the specific value of the hidden variable (i.e.\ position) in each experiment. Short of the above oracle,
Bohmian mechanics by itself is not only unable to predict the outcome of individual experiments, but  cannot even reproduce their provable (algorithmic) randomness properties (which follow from the Born rule). 
Furthermore, Bohmian mechanics cannot be extended by any deterministic theory so as to replace the random oracle, even if it is deemed acceptable that a deterministic theory fails to specify its initial conditions. 
Thus the performance of  Bohmian mechanics and especially its level of determinism are  similar to that of minimal versions of the Copenhagen interpretation, which also leaves the outcomes of  experiments to a black box (arguably a more obscure one). 

Partly to set the stage and partly to briefly discuss other aspects of determinism I will not go into here any further, let me summarize the formalism of Bohmian mechanics.\footnote{For details see e.g.\ 
Holland (1993), D\"{u}rr  \& Teufel (2009), Bricmont (2016),  and  Goldstein (2021).}

For simplicity, consider a system of $N$ non-relativistic spinless particles with masses $m_1, \ldots, m_N$, moving in $\R^3$, so that the classical configuration space is $\R^{3N}$. Suppose the classical Hamiltonian, defined on 
the associated phase space $T^*(\R^{3N})\cong\R^{6N}$, is given by 
\begin{align}
h(p,q)=\sum_{k=1}^N\frac{p_k^2}{2m_k} +V(q); && p=(p_1,\ldots, p_N), \:\: q=(q_1,\ldots, q_N).
\end{align}
The associated quantum theory starts from the Hilbert space $H=L^2(\R^{3N})$, on which one tries to define a Hamiltonian
$\hat{h}=-\sum_k \hat{\Dl}_k/(2m_k) +\hat{V}$ as a self-adjoint operator, where $ \hat{\Dl}_k$ is the Laplacian for the $k$th coordinate $x_k=(x_k^1,x_k^2,x_k^3)$ in $x=(x_1, \ldots, x_N)$, and 
  $\hat{V}$ is the potential  seen as a multiplication operator.  If this is done successfully,\footnote{In practice, $\hat{h}$ is first defined on some safe domain like $D=\cci(\R^{3N})$ and subsequently extended to a self-adjoint operator, but this procedure may be  unique, non-unique, or even impossible, depending on the potential $V$ (and, using configuration spaces $Q$ different from $\R^{3N}$, on $Q$). In the non-unique case (where $\hat{h}$ has several self-adjoint extensions) one faces some kind of indeterminism (Earman, 2009); interesting examples of this phenomenon come from spacetime singularities (Horowitz \& Marolf, 1995).
Such ``indeterminism''  will be inherited by Bohmian mechanics. But I will  ignore this on the understanding that situations where this happens can typically be resolved by providing \emph{classical} boundary conditions.}   by Stone's theorem any $\Psi\in H$ evolves unitarily for all $t\in\R$ as a solution of the Schr\"{o}dinger equation 
\begin{equation}
\hat{h}\Psi_t=i\hbar \frac{\partial \Psi_t}{\partial t}, \label{Sch}
\end{equation}
where $\Psi_0=\Psi$.
This solution, in turn, gives a probability measure
$P_t$ on $\R^{3N}$ via 
\begin{align}
dP_t(x)= \rh_t(x)d^{3N}x; &&
\rh_t(x):=|\Psi_t(x)|^2,  \label{mu}
\end{align}
where $d^{3N}x:=d^3x_1\cdots d^3 x_N$ with $d^3x_k=dx_k^1dx_k^2dx_k^3$. Since Born (1926), the role of $P_t$ is to provide probabilities for outcomes of measurements; e.g.\ if position is measured, then the probability  of finding the collective $N$-particle position $q$ in $A\subset\R^{3N}$ at time $t$ equals 
\begin{equation}
\mathrm{Pr}_t(q\in A)=P_t(A).\label{P}
\end{equation}
In textbook (Copenhagen) \qm\ these probabilities do not have an ignorance interpretation (and hence are ``irreducible''), because the particle positions are not just  \emph{unknown} before they are measured; they \emph{do not even exist} prior to measurement.

Bohmian mechanics may be seen as a modification of either classical or \qm. In the former view it adds 
 a (``pilot'') wave function $\Psi_t(q)$ to the particle trajectories $t\mapsto q(t)$,  whereas in the latter view it adds  particle trajectories to the wave function. The wave function is supposed to satisfy the  Schr\"{o}dinger equation of \qm, seen however as a classical \textsc{pde} (as opposed to an \textsc{ode} in \Hs).\footnote{As pointed out by 
Arageorgis \& Earman (2017), in Bohmian mechanics the wave function $\Psi$ is taken ``on the nose'', whereas in \qm\ it is its equivalence class in $L^2$ that defines the state.} But the wave function  only acts as a servant to the particles, in two completely different ways:
\begin{enumerate}
\item Once $\Psi_t(q)$ is known by solving \er{Sch}, the trajectories $t\mapsto q(t)$ are found by solving
\begin{align}
\frac{dq_k(t)}{dt}=v_k(q(t),t);&&
v_k(q(t),t):= 
\frac{\hbar}{m_k}\mathrm{Im}\left(\frac{\nabla_k\Psi_t(q(t))}{\Psi_t(q(t))}\right)=\frac{j_k(t,q(t))}{\rh_t(q(t))},
\label{Q}
\end{align}
where $j_k=\rh v_k$ is the usual $k$th particle probability current  of \qm, 
\begin{equation}
j_k(t,x)=\frac{i\hbar}{2m_k}(\ovl{\Psi_t(x)}\,\nabla_k\Psi_t(x) - \Psi_t(x) \nabla_k\ovl{\Psi_t(x)}).
\end{equation}
Here $\nabla_k=(\p/\p x^1_k, \p/\p x^2_k, \p/\p x^3_k)$. Provided $\Psi_t$ satisfies Schr\"{o}dinger equation \er{Sch}, the current $j_k$ is related to the probability density \er{mu} by the conservation law
\beq
\p_t\rh_t(x)+\Sigma_k \nabla_k\cdot j_k(t,x)=0.
\eeq
Though \er{Q} is just a flow equation for a time-dependent vector field $v$,
 \er{Sch} - \er{Q} form a coupled \textsc{pde-ode} system of a novel kind that has no counterpart in either classical or \qm. Although Stone's theorem still applies to \er{Sch}, the ensuing vector field $v$ may be incomplete, in which case the trajectories $q(t)$ do not exist for all time and all initial conditions. This adds a potential source of indeterminism to Bohmian mechanics that has no analogue in \qm\ but is unrelated to  the ``random-ish'' kind of indeterminism discussed  in this paper.\footnote{For details see Berndl \emph{et al}.\ (1995), Teufel \& Tumulka (2005), and Arageorgis \& Earman (2017).
}
\item Though considered real, the particle locations $q(t)$ are mysteriously hidden \emph{in principle} and can only be revealed by measurement.  All we can know is their probability distribution, which at time $t$ is given by the ``quantum equilibrium measure'' \er{mu} - \er{P}, just like in \qm.  This move turns Bohmian mechanics into a special case of classical statistical mechanics, albeit
with the unusual features that:
\begin{enumerate}
\item The probabilities are defined on configuration space (rather than phase space).
\item A consistent interpretation of these probabilities is elusive. On the one hand, the assumed reality of the particle positions and trajectories forces an ignorance interpretation: since at any time $t$ all particles are supposed to actually \emph{have} positions, $\rh_t$ (which never describes these actual values) cannot be objective. But on the other hand, since these values cannot even be known in principle (until they are measured), the probabilities defy an ignorance interpretation and seem to be ontic or objective. See also Myrvold (2021), \S 9.4. This dilemma haunts Bohmian mechanics and may explain some of its ultimate indeterminism.
\end{enumerate}
\end{enumerate}
In sum, although its role as a ``pilot'' in 1.\ gives the wave function $\Psi_t$  some ontic status, its associated probability distribution $\rh_t$ uncomfortably floats between subjectivity and objectivity. According to
 D\"{u}rr,  Goldstein, \& Zangh\`{\i} (1992), $\rh_t$  determines which particle configurations are deemed ``typical'' in the sense of  \emph{``overwhelmingly probable''}, but this shifts the burden of interpretation to the question what \emph{that} is supposed to mean! See \S\ref{discussion}.
 \section{What makes a hidden variable theory deterministic?}\label{S2}
 Since my analysis does not only apply to Bohmian mechanics and may perhaps even be clearer in a more general context, I momentarily take a broader perspective. 
A general \hv\ theory $T$ underneath \qm, say in the specific context of some given well-understood and repeatable experimental setting, should provide at least:\footnote{See Leifer (2014) for a  more detailed analysis of a setting like this. See also Bub (2022) for history.}
\begin{itemize}
\item A space $\Lm$ of hidden variables $\lm$; in Bohmian $N$-particle mechanics this is  position space $\Lm=\R^{3N}$, so that $\lm=q=(q_1,\ldots, q_N)$ with $q_k=(q_k^1, q_k^2,q_k^3)$, as above.
\item An assignment $\Psi\mapsto P_{\Psi}$, where $\Psi$ is a quantum-mechanical state (not necessarily a wave function) describing the system, and $P_{\Psi}$ is a  probability measure  on $\Lm$. In Bohmian mechanics  $P_{\Psi}$ is the Born/``equilibrium'' measure \er{mu} at time $t$.
\item Conditional probabilities $P_{\lm}(e\mid a)$ stating the probability of outcome $e$ of the experiment on setting $a$, given the value $\lm$ of the \hv; in Bohmian mechanics this of course depends on the details of the experiment, which I need not go into.
\end{itemize}
The  compatibility requirement between \qm\ and the given theory, then, is
\begin{equation}
\mathsf{P}_{\Psi}(e\mid a) =\int_{\Lm} dP_{\Psi}(\lm)\, P_{\lm}(e\mid a),\label{Bell}
\end{equation}
where $\mathsf{P}_{\Psi}(e| a)$ is the quantum-mechanical Born probability for outcome $e$ given $a$ and $\Psi$.
In this formalism, a \hv\ theory $T$ is  called \emph{deterministic} (just terminology!)  if 
\begin{equation}
P_{\lm}(e\mid a) \in\{0,1\} \mbox{ for any value of } \lm, \, e,\, a. \label{ND}
\end{equation}
That is, the value of $\lm$ \emph{determines} the outcome of the experiment.\footnote{In Bohmian mechanics his
 is true if one knows the hidden positions exactly at the time of measurement. At earlier times, the pilot wave = quantum state $\Psi$ is needed to make predictions, since $\Psi$ determines the trajectories. See e.g.\ Barrett (1999), Chapter 5, for a  discussion of measurement in Bohmian mechanics.} On this basis, exploring the  bipartite (``Alice and Bob'') setting introduced by Einstein, Podolsky, and Rosen (1935), Bell (1964)  proved that the conjunction of the following properties is inconsistent:
\begin{enumerate}
\item \emph{Determinism} in the (narrow) sense of \er{ND} and preceding definitions;
\item  \emph{Quantum mechanics}, i.e.\  the compatibility rule \er{Bell};
\item \emph{Locality} in a sense made precise in Bell (1964);
\item  \emph{Free choice}, i.e.\ (statistical) independence of  Alice's and Bob's settings $a$ and $b$ from each other and from the \hv\ $\lm$ (given the probability measure $P_{\Psi}$). 
\end{enumerate}
There are  four (minimal) ways out of this contradiction: just reject one of the assumptions! 
\begin{enumerate}
\item The Copenhagen interpretation rejects determinism (I take this interpretation to be a ``family resemblance''
but all versions should agree on this point).
\item 
In arguing that the Born distribution \er{mu} should be the end result of an equilibration process and 
hence did not always hold in the past,
 Valentini (1992, 2020) effectively departs from the Born rule and hence \qm. See also \S\ref{discussion}.
 \item    Bohmian mechanics is non-local (D\"{u}rr \& Teufel, 2009, Chapter 10; Goldstein, 2021, \S 13). The question if this non-locality was really necessary motivated Bell (1964).
 \item Finally, 
 't Hooft (2016) rejects free choice in his deterministic cellular automaton interpretation of \qm.\footnote{Bohmians do not explain the origin of 
 the  settings of the experiment, which are simply left out of the theory. In my view this weakens their case for determinism even further, but this is not my main point.} See also Hossenfelder \& Palmer (2020).
 \end{enumerate}
Restoring determinism in the quantum world is not the only goal, perhaps not even the most important goal of \hv\ theories. This applies to  Einstein, and Bell (after whom I will quote similar words from Bricmont) continues the  quotation in \S\ref{intro} with:
\begin{quote}
\begin{small}
More importantly, in my opinion, the subjectivity of the orthodox version, the necessary reference to the ``observer'', could be eliminated. (Bell, 1982, p.\ 990).

While deterministic, the de Broglie--Bohm theory also accounts naturally for the apparent indeterminism of quantum phenomena.  
Finally, it explains the ``active role'' of measuring devices (\ldots), so strongly emphasized by the Copenhagen school, but making it a consequence of the theory and not of some a priori philosophical doctrine. It also explains the nonlocal interactions inherent in quantum phenomena. What more could we ask for?\hfill (Bricmont, 2016, p.\ 19).
\end{small} 
\end{quote}
Having said this,  much as one may read Bell (1964) both as a proof that \qm\ is non-local in a suitable sense \emph{and} as a proof that deterministic \hv\ theories (in the first sense defined below) are necessarily non-local (in a slightly different sense),\footnote{It is a miracle to me why the Bohmians so vehemently oppose the second reading (Bricmont, 2022).} one may also look at Bohmian mechanics as an attempt to provide a realist ontology for \qm, \emph{and} as an attempt to restore determinism.\footnote{Instead of as a \hv\ theory, Bohmian mechanics may also be seen as an observer-free no-collapse interpretation of \qm, see Bub (1997). Here the wave function $\Psi$ is not so much seen as a ``real'' physical field (with mathematically of course is complex!) comparable with the electromagnetic field, but as a mathematical device that controls the modal properties of the theory.} These goals are closely related, since they are most easily achieved simultaneously by rewriting \qm\ as a classical theory, which is what Bohmian mechanics essentially tries to do.

However, the definition of determinism used above, which is typical for the \hv\ literature including Bohmian mechanics, strikes me as too weak, since nothing is said about individual experiments. Recall that for Born (1926), collision theory marked a fundamental difference between classical and \qm\ in that: \emph{the initial conditions being known in both cases}, the former could predict the outcome with certainty whereas the latter could not.
 Classical physics may be said to be ``indeterministic'' on account of its typical inability to provide or explain initial conditions (say for the solar system),  but this is not the kind of indeterminism that concerned Born, or me. 
I will argue that in Bohmian mechanics, \emph{even if the initial conditions (of either the universe or some relevant part thereof), i.e.\  the exact particle positions, were known}, no deterministic account of the very specific and provable random structure of typical outcome sequences of quantum-mechanical experiments can be given. To this end I will show that in order to actually acquire the mathematical information necessary to predict such outcome sequences the theory requires an external oracle \emph{whose functioning cannot even in principle be described by any deterministic theory, including of course Bohmian mechanics itself}. 

In particular, the randomness of these quantum-mechanical outcome sequences cannot be a consequences of whatever randomness in the initial condiitons, as suggested by  D\"{u}rr,  Goldstein, \& Zangh\`{\i} (1992); see \S\ref{discussion}.
Let me now try to make this argument more precise. 
  \section{Is Bohmian mechanics deterministic?}\label{S3}
First,  as explained in Landsman (2020, 2021), in the narrow context of \emph{deterministic} \hv\ theories, refined mathematical methods are available (namely from the theory of algorithmic randomness) that make an appeal to a  bipartite setting unnecessary: since my argument does not rely on entanglement, we may  simply work with a quantum coin toss (realized for example as a spin-$z$ measurement on a spin-$\half$ particle in the state $(1,1)/\sqrt{2}$, or optically using suitable polarizers).
 The settings, possible contexts, and quantum state of the experiments are then fixed to be the same for each experiment in a long run,
 so that only the hidden state $\lm$ (i.e.\ in Bohmian mechanics the particle positions $q$) may change. Idealizing to an infinite run, one then simply has an outcome sequence $$s:\N\raw\{0,1\},$$ such that $e_n=s(n)$ is the outcome of experiment no.\ $n$ (with the given settings etc.).
 
Although standard (Copenhagen) \qm\ refuses to say anything about the origin of each outcome $s(n)$,  it does make very specific statistical predictions. These do not just concern the single-case probabilities for individual outcomes, which are given by the Born rule (and are 50-50 in the case at hand), but include far more refined claims  \emph{about the sequence $s$ as a whole}, which go well beyond saying that the probability of each outcome is 1/2. The basis of these more detailed predictions is the following theorem:\footnote{This is 
 Corollary 3.4.2 in Landsman (2021), which is  Corollary 4.2 of the version \url{arXiv:2003.03554}.}
   \begin{theorem}\label{keycor}
  With respect to the standard ``fair'' Bernoulli probability measure on the space $\ul{2}^{\N}$ of all binary sequences,\footnote{This measure is the extension of the 50-50  probability for single bits to sequences (Dudely, 1989, \S 8.2).} almost every outcome sequence $s$ of an infinitely often repeated fair quantum coin toss is Kolmogorov--Levin--Chaitin random (i.e.\  1-random).\footnote{See Calude (2002) or Downey \& Hirschfeldt (2010) for complete treatments, and
Appendix B of Landsman (2021) for a quick summary. Dasgupta (2011) is intermediate between these extremes. This concept was first defined by Kolmogorov  for binary \emph{strings} (which are finite by convention) and was subsequently extended to  (infinite) binary \emph{sequences} by Levin and independently by Chaitin; whence the name `Kolmogorov--Levin--Chaitin' randomness.
The `1' in the more technical name `1-randomness' refers to a family of similar notions (Downey \& Hirschfeldt, 2010) whose use in physics should also be explored. Very briefly, and roughly: the  \emph{Kolmogorov complexity} $K(\sg)$ of a binary \emph{string} $\sg$ is the length of ``the'' shortest computer program $P$ that prints $\sg$ (and then halts). Then $\sg$ is 
 \emph{Kolmogorov random} if $K(\sg)=|\sg|+ O(\log|\sg|)$, which means that
\emph{the shortest computable description of  $\sg$ is $\sg$ itself}. 
 One may think of this as $K(\sg)\approx |\sg|$. A binary \emph{sequence}  $x$ is \emph{Kolmogorov--Levin--Chaitin random} if
 there exists $c\in\N$ such that $K(x_{|N})\geq N - c$ for each $N\in\N$, where $x_{|N}$ is the truncation of $x$ to its first $N$ bits. This definition justifies using infinite sequences as idealizations of long finite strings (where ``long'' means: $N \gg c$ for given $x$), in so far as randomness properties are concerned. I will freely do so in what follows. \label{long}
} \end{theorem}
This  notion of randomness (in this case of binary sequences) arose in the 1960s when ideas from probability theory were combined with the theory of algorithms and computation (\`{a} la Turing).  I already motivated its use in quantum theory (and elsewhere) in Landsman (2021, 2022) and will return to this motivation in \S\ref{discussion} below in connection with the notion of typicality used by the Bohmians.  But here I use this specific concept of randomness because Theorem \ref{keycor} is simply a theorem, whose relevance will become clear shortly.
 
I will  show that Theorem \ref{keycor} leads to insurmountable tension with determinism. First, note that in ``deterministic''
 \hv\ theories the outcomes $s$ factor through $\Lm$, i.e.,
 \begin{center}
  there are functions $h:\N\raw\Lm$ and $g:\Lm\raw\{0,1\}$ such that 
  \beq
  s=g\circ h.\eeq 
  \end{center}
\begin{itemize}
\item   The existence of $g$ expresses the idea that  the value of $\lm$ \emph{determines} the outcome of the experiment, cf.\ \er{ND}, which is \emph{one} important sense in which hidden variable theories could try to be deterministic. The function $g$ incorporates all details of the experiment that may affect its outcome, like the setting $a$,  a possible context $C$, and the quantum state $\Psi$,  \emph{except the hidden variable $\lm$, which $g$ assumes as its argument}.\footnote{In more general measurements $g$ takes values in the spectrum of the operator that is measured.}   In this light, it would be more precise to write 
$g(\lm)=g_{(a,C,\Psi)}(\lm)$,
  but since we agreed that $(a,C,\Psi)$ and whatever else is relevant for the outcome sequence is fixed throughout the entire run, for simplicity I just write $g(\lm)$. In Bohmian mechanics the contextuality of $g(q)$ 
via its hidden dependence on $a$ and $C$ is well known (e.g.\ D\"{u}rr \& Teufel, 2009, \S 12.2.3; Goldstein, 2021, \S 12). Answering a F.A.Q.,
the existence of $g$ therefore does not contradict the Kochen--Specker theorem.  
  \item
 The function $h$ gives the value of $\lm$ in experiment number $n$ in a long run, for each $n$. Conceptually, 
 $h$ samples the probability measure $P_{\Psi}$ relevant to the quantum-mechanical state in which the experiment is prepared (see  \S\ref{discussion} for further discussion).
 For example, in Bohmian mechanics $h:\N\raw\R^{3N}$ in principle picks an element 
 \beq
 h(n)\equiv q(n)=(q_1(n),\ldots, q_N(n))\in\R^{3N},
 \eeq
i.e., an $N$-particle configuration in $\R^3$ (seen as the total space of the universe, assumed to contain $N$ particles), for each $n\in\N$. This is vast, but as I will review in  \S\ref{discussion},  under suitable independence and preparation assumptions  the results of D\"{u}rr,  Goldstein, \& Zangh\`{\i} (1992) enable one to replace
$h:\N\raw\R^{3N}$ by $h':\N\raw\R^{3}$.
\end{itemize}
   \begin{theorem}\label{TAB}
 The sampling function  $h$ cannot be provided by Bohmian mechanics, or by some hitherto unknown extension of it that may reasonably be claimed to be deterministic. 
   \end{theorem}
\emph{Proof.}  Arguing by \emph{reductio ad absurdum}, if $h$ is provided by some deterministic theory $T$, right because $T$ is supposed to be deterministic,  $h$ explicitly gives the values $x_n=h(n)$ for each $n\in\N$ (i.e.\ experiment no.\ $n$). 
Since $g$ is also given, this means that the sequence $s=g\circ h$ is described explicitly by some formula. By Chaitin's second incompleteness theorem,\footnote{This is Theorem B.4 in Landsman (2021), both in the printed and the arXiv version. Briefly, this theorem states that  even   comprehensive mathematical theories (such as set theory) can compute only finitely many digits of a 1-random sequence $x$. See e.g.\ Calude (2002), Theorem 8.7, which is stated for Chaitin's famous random number $\Omega$ but whose proof holds for any 1-random sequence. In view of Theorem B.1 (\emph{loc.\ cit.}), also due to Chaitin, this argument applies equally well to long strings (cf.\ footnote \ref{long}).
} the outcome sequence cannot then be 1-random, against Theorem \ref{keycor}.\hfill Q.E.D.
 \medskip
 
Thus the question arises \emph{what else provides $h$}. In order to recover the predictions of \qm\ as meant in Theorem \ref{keycor}, i.e.\ not just the single-case Born probabilities but the (Kolmogorov--Levin--Chaitin) randomness properties of entire outcome sequences,  the function $h$ must sample either the Born measure $P_{\Psi}$ on $\Lm=\R^{3N}$, seen as the ``quantum equilibrium distribution'' of the universe, or 
the one-particle Born measure on $\R^3$ provided by some effective wave function $\psi$ (see above). Since $g$ is  supposed to be given by Bohmian mechanics, this implies that the randomness properties of $s$ must entirely originate in $h$.   
  This  origin cannot be deterministic, since in that case we are back to the contradictory scenario  above. Hence $h$ must come from, or indeed may be identified with, some unknown external random process in nature that Bohmian mechanics needs to invoke as a kind of random oracle--it cannot be overemphasized how strong the requirement of \emph{random sampling} of the configuration space $\R^{3N}$ with respect to the probability measure $P_{\Psi}$ is; as shown above, this requirement excludes determinism (and, if different, computability).
   \section{Discussion}\label{discussion}   
Theorem \ref{TAB} allows a sharp  comparison with Copenhagen \qm:
\begin{itemize}
\item In Bohmian mechanics  the outcome sequence $s=g\circ h$ factors though its hidden variable space $\R^{3(N)}$ 
and the corresponding sampling function $h:\N\raw\R^{3(N)}$ is its source of randomness, lying outside the theory as a black box or external ``oracle''.
\item  In Copenhagen quantum mechanics there is no such  factorization of $s$; here, the  unanalyzed black box and source of randomness is the
measurement process itself.
\end{itemize}
So although in Bohmian mechanics  the source of indeterminism has been shifted compared to its place in Copenhagen quantum mechanics, this source
 certainly has not been removed.
 
 As especially stressed by  D\"{u}rr,  Goldstein, \& Zangh\`{\i} (1992), D\"{u}rr \& Teufel (2009), and Goldstein (2021), Bohmian mechanics  tries (or: Bohmians try) to answer all such worries by an appeal to \emph{typicality}. To this end, their main weapon is what they call the \emph{quantum equilibrium hypothesis}, which I will now try to summarize.\footnote{I try to follow D\"{u}rr \& Teufel (2009), Chapter 11, who however omit the phase factor in \er{4.6}. See also Norsen (2018), \S 5, whose derivation should be adjusted in the same way, and Myrvold (2021), \S 9.4.}
  In the analysis reviewed in \S\ref{intro} we take $\Psi$
to be the wave function of the universe, seen as a system of $N$ particles moving in $\R^3$, where $N$ is very large (like $N=10^{80}$). We now take $M$ large enough to describe the number of measurements in a long run, but small compared to $N$ (think of $M=10^{10}$), and accordingly split $x=(y,z)$ with $y=(x_1,\ldots x_M)$ and $(z=x_{M+1}, \ldots, x_N)$. Similarly, we write $q=(r,s)$ for the actual particle positions.\footnote{Seen as random variables on the probability space $\R^{3N}$, $q$ consists of the coordinate functions $q^i_k(x)=x_k^i$ for $k=1, \ldots, N$ and $i=1,2,3$, and similarly
 $r^i_k(y,z)=y_k^i$ for $k=1, \ldots, M$ and $i=1,2,3$, etc.} The \emph{conditional wave function} 
\begin{align}
\Psi^s(y)=\frac{\Psi(y,s)}{\|\Psi(\cdot,s)\|_{L^2(\R^{3M})}}; && \|\Psi(\cdot,s)\|_{L^2(\R^{3M})}=\sqrt{\int d^{3M}y\, |\Psi(y,s)|^2}
\end{align}
of the $M$-particle system then trivially gives the  conditional Born probabilities for $r$, i.e.\footnote{This is
really a shorthand for $P_{\Psi}(r\in A\mid s)=P_{\Psi^s}(A)=\int_A d^{3M}y\, |\Psi^s(y)|^2$, where $A\subset\R^{3M}$.}
\begin{equation}
dP_{\Psi}(r\mid s)=dP_{\Psi^s}(r).
\end{equation}
Now take some fixed wave function $\psi_M\in L^2(\R^{3M})$, i.e.\  $\psi$ is a function of $y\in \R^{3M}$, where 
\beq
\|\psi_M\|_{L^2(\R^{3M})}=1,
\eeq
and assume that the ``environmental'' particle positions $s\in \R^{3(N-M)}$ decouple in that
 \beq
\Psi(y,s)=\psi_M(y)\Phi(s),\label{fac} \eeq 
for some $\Phi\in  L^2(\R^{3(N-M)})$,  whilst the $r$ variables have autonomous Bohmian dynamics: $\psi_M$ should satisfy a Schr\"{o}dinger equation with a potential $V(y)$ not containing $z$, and the flow equations \er{Q} for $r$ do not contain $s$.  The factorization \er{fac} then
lasts as long as we use $\psi_M$ (i.e.\ during the repeated measurements to be described below), 
and we have
\begin{align}
\Psi^s(y)=e^{i\al(s)}\psi_M(y); && e^{i\al(s)}=\frac{\Phi(s)}{|\Phi(s)|}.
\end{align}
Thus $\exp(i\al(s))$ is a phase factor.\footnote{Like in \qm, also in Bohmian mechanics wave functions differing by a phase should be identified, since both the velocity field \er{Q} and the probability distribution \er{mu} are the same.}  Hence for any $y\in\R^{3M}$, $(y,s)\in\R^{3N}$ lies in the set
\beq
\{\Psi\cong\psi_M\}=\{(y,z)\in \R^{3N}\mid\Psi^z(y)=e^{i\al(z)}\psi_M(y)\mbox{ for some } \al:\R^{3(N-M)}\raw\R\}.
\label{4.6}
\eeq
We may finally ``trace out'' (i.e.\ marginalize) over all such $s$ to arrive a the 
formula
 \begin{equation}
dP_{\Psi}(r\mid \Psi\cong\psi_M)=dP_{\ps_M}(r).
\end{equation}
If we finally assume that $\ps_M(y)=\psi(y_1)\cdots \psi(y_M)$ for some normalized $\psi\in L^2(\R^3)$, then
  \begin{equation}
P_{\Psi}(r\in A\mid \Psi\cong\Pi_{k=1}^M\psi_k)=\int_A d^{3M}y\, |\psi(y_1)|^2 \cdots |\psi(y_M)|^2,\L{QEH} 
\end{equation}
for any (measurable) $A\subset \R^{3M}$, where $\psi_k(y_k)=\psi(y_k)$ for the given single-particle wave function $\psi$. This means that under the stated independence assumptions (which one should be able to satisfy in setting up a repeated single-particle measurement whose quantum-mechanical description involves identical initial conditions for each experiment but whose Bohmian description allows for differences in the particle locations $r$ in each case), the probability distributions of the given $M$-particle system induced by the quantum equilibrium distribution for the wave function of the universe $\Psi$ equals the Born  distribution.
Eq.\ \er{QEH}  is what  D\"{u}rr,  Goldstein, \& Zangh\`{\i} (1992) call the \emph{quantum equilibrium hypothesis}, which, then, is actually a theorem (albeit under  restrictive independence assumptions):\footnote{It would appear to be more natural to me to use the name ``quantum equilibrium hypothesis'' to the postulate that the $N$ particles comprising the material content of the universe $\R^3$ were, and hence are, distributed according to the Born probabilities $|\Psi|^2$, where $\Psi$ is the wave function of the universe.}
\begin{quote}
\begin{small}
We prove that for \emph{every} initial $\Psi$, this agreement with the predictions
of the quantum formalism is obtained for \emph{typical}---i.e., for the overwhelming
majority of---choices of initial $q$. And the sense of typicality here
is with respect to the only mathematically natural---because equivariant---candidate at hand, namely, quantum equilibrium. Thus, on the universal level, the physical significance of quantum
equilibrium is as a measure of typicality, and the ultimate justification of
the quantum equilibrium hypothesis is, as we shall show, in terms of the
statistical behavior arising from a typical initial configuration. 
\hfill (D\"{u}rr,  Goldstein, \& Zangh\`{\i}, 1992, p.\ 858)
\end{small} 
\end{quote}
While \er{QEH} does support the Bohmian mechanics formalism and relates it to experiment, 
it does not materially affect my discussion in \S\ref{S3}. In that context, all the quantum equilibrium hypothesis shows is that in the Bohmian description of repeated experiments based on identical single-particle states with wave function $\psi$, instead of sampling from the probability distribution of all particles in the universe given by $|\Psi|^2$ one may sample from the single-particle  probability distribution $|\psi|^2$. But this leads to exactly the same problems. 

Restricting ourselves to quantum coin tosses, as before, and assuming with the Bohmians that every measurement can ultimately be reduced to position measurements, eq.\  \er{QEH} is simply replaced by the 50-50 Born probabilities used in \S\ref{S3}. The quantum equilibrium hypothesis/theorem then states that, provided the initial particle configuration of the universe is typical, so is the 
outcome sequence of an endlessly repeated quantum coin toss. This matches the ``almost every'' in the statement of Theorem \ref{keycor}, and in this experiment we may therefore identify the ``typicality'' of an outcome sequence with its 1-randomness.

The idea that, at least in the context of Theorem  \ref{keycor} (and, I believe, also in the context of classical statistical physics),\footnote{For a start, see Gr\"{u}nwald \& Vit\'{a}nyi (2003). What is lacking so far is, among others, the connection between algorithmic randomness and large deviation theory (Ellis, 1985; Touchette, 2009).}
 1-randomness (=  Kolomogorov--Levin--Chaitin randomness)
 is the correct notion of typicality for binary sequences has been argued by Landsman (2021, 2022), and should be uncontroversial. It is much more \emph{precise} than the notion of ``overwhelmingly probable'' used e.g.\ in  D\"{u}rr,  Goldstein, \& Zangh\`{\i} (1992),
 D\"{u}rr \& Teufel (2009), and elsewhere, and is much \emph{stronger} than (Borel) normality,\footnote{A sequence $s$ is \emph{Borel normal} if each possible finite string $\sg$ in $s$ has (asymptotic) frequency $10^{-|\sg|}$ 
    (so that each digit $0, \ldots, 9$ occurs 10\% of the time,  each block $00$ to $99$ occurs 1\% of the time, etc.).}
 which is also sometimes used in the literature on Bohmian mechanics (e.g.\ Callender, 2007). However, Borel normal sequences may be computable,\footnote{In base 10 the simplest example is   \emph{Champernowne's number}
 $01234567891011121314151617181920 \ldots$, which
 can be shown to be Borel normal. The decimal expansion of $\pi$ is conjectured to be Borel normal.}
  and hence despite their abundance and  attractive statistical properties they are useless in the context of \qm, whose randomness properties are much stronger than normality, as Theorem \ref{keycor} shows.\footnote{Any
 1-random sequence  is Borel normal (Calude, 2002, Corollary 6.32).}

The identification of ``typicality'' with ``overwhelming probability'' is always predicated on a background measure, which needs to be justified. For a repeated quantum (or even  classical) coin flip this background measure declares that each binary string (of some given length) or sequence is equally likely to occur. In particular, 
 a sequence like $10110000101111$ is as probable as $00000000000000$. The intuition that the former is typical in a probabilistic sense while the latter is not, applies to the coarse-graining obtained by counting zeros and ones; indeed, 
 $(6,8)$ (i.e.\ a sequence with 6 zeros and 8 ones) is typical whereas $(14,0)$ is not. The notion of 1-randomness captures this difference already at the level of the sequences themselves, without any need for coarse-graining;
 though it occurs with probability one on the space of binary sequences, the concept of 1-randomness is not itself defined probabilistically (but algorithmically). What seems lacking in Bohmian mechanics is a similar concept of typicality of $N$-particle configurations that is defined intrinsically, and subsequently can be shown to occur  with probability one relative to the background measure provided by
 the equilibrium distribution $|\Psi|^2$ defined by the wave function of the universe $\Psi$. Defining typicality \emph{by} this background measure would, in the absence of coarse graining, be similar to saying that all binary strings or sequences are typical---for they all have the same probability relative to the pertinent background measure, as just pointed out. This makes it hard to understand what could be meant by `\emph{typical} choices of $q$' (in the above quotation); note once again that appealing to an `overwhelming majority' only makes sense after coarse graining. In this light, consider the Bohmian ideology that:
\begin{quote}
\begin{small}
For a universe governed by
Bohmian mechanics (\ldots), given the initial wave function
and the initial positions of all particles, \emph{everything} is completely determined
and nothing whatsoever is actually random. (\ldots)
The origin of the randomness in the results of quantum measurements lies in random initial
conditions, in our ignorance of the complete description of the system
of interest---including the apparatus---of which we know only the wave
function. \hfill (D\"{u}rr,  Goldstein, \& Zangh\`{\i}, 1992, pp.\ 846, 844)
\end{small} 
\end{quote}
In my view, this ideology rests on an equivocation in which the initial conditions of the universe (simplified as some $N$-particle configuration in $\R^{3N}$) are simultaneously treated as uniquely given and hence singular (though inaccessible in principle) and yet  subject to some probability distribution with respect to which they are supposed to be typical.\footnote{To make things  worse, this distribution is also unknown in practice, since its source $\Psi$ is unknown.}

 Random initial conditions are drawn from a sample space. The (unique) result of this draw induces a binary outcome string or sequence $h$ for a repeated quantum coin toss for which some specific  $M$-particle subsystem has been used.
Choosing a different $M$-particle subsystem would probably give a different outcome sequence $h'$, but in Bohmian mechanics both are ultimately \emph{determined} by the initial $N$-particle configuration. Yet according to \qm, as mimicked by Bohmian mechanics via \er{QEH} (and more generally by its aim to preserve all statistical results of \qm), \emph{each} of these sequences is 1-random \emph{by itself}, cf.\ Theorem \ref{keycor}. By  Theorem \ref{TAB} this randomness  cannot be a consequence of the randomness of the initial conditions, since in Bohmian mechanics these are \emph{fixed} and everything afterwards is  supposed to be deterministic. In other words, the above claim that `The origin of the randomness in the results of quantum measurements lies in random initial conditions' cannot be upheld, since there is no way for the ``randomness'' in  initial conditions to influence measurement results $h$ and hence ``cause'' \emph{their} randomness.\footnote{I am grateful to Joanna Luc for questions in this direction at the Cambridge--LSE PoP Bootcamp.}

In comparison, take classical coin tossing. Here, the initial conditions are (to a good approximation) the vertical velocity $v$ and rate of spin $\om$ of the coin at the time it is launched; see Diaconis, Holmes, \& Montgomery (2007) or Diaconis \& Skyrms (2018). This time, the (apparent) randomness in an outcome sequence is genuinely ``caused'' by randomness in the initial conditions, which are sampled by varying wrist and thumb movements, and hence ultimately by some sort of brain process. Due to the extreme sensitivity of the outcome to the initial conditions, 
almost any non-sharply peaked probability distribution on the space of initial conditions leads to a 50-50 distribution on the binary outcome space.\footnote{In fact, there is a slight 1\% bias for the coin to land the way it started (see references in main text). A Galton board provides a similar example with a richer outcome space (D\"{u}rr \& Teufel,  2009, \S 4.1.1).}

 Nonetheless, as in the case of \qm\ (see Theorems \ref{keycor} and \ref{TAB}) the outcome sequence would only be 1-random if the ultimate sampling process in the brain were indeterministic (in which case it might be quantum mechanical). Indeed, by the same reasoning as in the proof of Theorem \ref{TAB}, any kind of determinism blocks 1-randomness. In particular, sequences produced by coin tossing machines are not 1-random and I would not be surprised if those produced by humans aren't either. But this is no problem, since few scientist would maintain that the probabilities in coin tossing are fundamental; in contrast, the Born probabilities \emph{are} regarded as exact and fundamental, also by the
 Bohmians.

This discussion suggests  a way out, however. Theorems \ref{keycor} and  \ref{TAB} assume the exact validity of the Born rule. But in my view no probability measure used in mathematical physics is ever exact, because the physical origin of the measure lies in the way it is sampled.\footnote{Boltzmann's program of justifying the microcanonical ensemble by ergodic theory (Uffink, 2007,  2022) is paradigmatic;  ``typical'' behaviour would be dynamically generated via time averages. The Bohmians  disagree with this goal (Lazarovici \& Reichert, 2015), but this is not the disagreement in the main text. 
} Thus I expect the origin of the Born measure to lie in some underlying theory from which \qm\ is emergent \`{a} la Butterfield (2011). This theory may well be deterministic---\emph{provided the Born rule is an approximation}. The program of Bohm \& Vigier (1954) and Valentini (1992, 2020), who try to \emph{derive}  quantum equilibrium  from equilibration, is  in this spirit (see also Norsen, 2018 and Myrvold, 2021), especially because it accepts departures from the Born rule (also cf.\ \S\ref{S2}).
The only disagreements I would have with  Valentini (2020) is that I would not describe the Bohmian
 arguments based on typicality as `circular' but as \emph{inconclusive} (see above), and that I would prefer
the underlying theory not to be the de Broglie-Bohm pilot-wave theory but something really new. Whatever this new theory may be, its sampling process should be 
washed out, as in the ``method of arbitrary functions'' in probability theory (Engel, 1992; Myrvold, 2021). 
Indeed, this is also what happens in  examples like  coin tossing (even if it is deterministic).
\begin{small}

\end{small}
\section*{Data availability statement}
The data for this article consist of the reasoning and the references. All references are easily found either on the internet or, in case of books,  in standard academic libraries. Readers who have any trouble finding a reference are welcome to contact the author.\end{document}